\documentclass[12pt]{article}
\usepackage[]{epsfig}

\begin{document}

Paper contributed to the 
X International Symposium on Lepton and Photon Interactions at High Energies,
 23rd-28th July 2001, Rome, Italy. 

{\LARGE \begin{center}
Results and plans of the CRESST dark matter search
\end{center}}

\begin{large}
\begin{center}
M.\,Altmann$^b$, 
G.\,Angloher$^d$, 
M.\,Bruckmayer$^b$, 
C.\,Bucci$^a$
S.\,Cooper$^d$, 
C.\,Cozzini$^b$, 
P.\,DiStefano$^b$, 
F.\,von Feilitzsch$^c$, 
T.\,Frank$^b$, 
D.\,Hauff$^b$, 
Th.\,Jagemann$^c$, 
J.\,Jochum$^c$,
R.\,Keeling$^d$, 
H.\,Kraus$^d$, 
J.\,Macallister$^d$, 
F.\,Pr\"obst$^{b,}$\footnote[2]{Corresponding author; Tel.: +49 89 32354 270; E-mail: proebst@mppmu.mpg.de}
Y.\,Ramachers$^d$,
J.\,Schnagl$^c$, 
W.\,Seidel$^b$,  
I.\,Sergeyev$^{b,1}$, 
M.\,Stark$^c$, 
L.\,Stodolsky$^b$,
H.\,Wulandari$^c$
\end{center}
\end{large}
\centerline{\it\small $^a$ Laboratori Nazionali del Gran Sasso,
I-67010 Assergi, Italy}
\centerline{\it\small $^b$ Max-Planck-Institut f\"ur Physik,
F\"ohringer Ring 6, D-80805 Munich, Germany}
\centerline{\it\small $^c$ Technische Universit\"at M\"unchen,
Physik Department, D-85747 Munich, Germany}
\centerline{\it\small $^d$ University of Oxford, Physics Department ,
Oxford OX1 3RH, U.K.}

\begin{abstract}
Data taken by CRESST in 2000 with a cryogenic detector system based on 262 g sapphire 
crystals is used to place limits on WIMP dark matter
in the Galactic Halo.  The detector is especially sensitive for
low-mass WIMPS with spin-dependent cross sections and improves 
on existing limits in this region.
CRESST is now preparing for a second phase,  which will use a 10 kg detector consisting of 300\,g CaWO$_4$ 
crystals with simultaneous detection of phonons and scintillation light
to reduce  background.
\end{abstract}
{\it \small $^1$ Permanent Address: Joint Institute for Nuclear
Research, Dubna, 141980, Russia}\\

\newpage

The goal of the CRESST\footnote{
Cryogenic Rare Event Search with Superconducting Thermometers}
experiment is to detect WIMP dark matter particles via the energy 
they deposit when elastically scattering on nuclei.
We have developed very sensitive massive cryogenic detectors
for this purpose and installed them in a low-background facility
in the Gran Sasso Underground Laboratory (LNGS).

The first section of this paper describes our low-background facility in LNGS.
The second section describes Phase I of the project,
which used 262\,g sapphire cryogenic calorimeters, 
and presents the resulting dark matter limits.
The final section describes our plans for Phase II,
using scintillating calorimeters to reduce the background.

\section{Low-background facility}

\begin{figure}[p]
\vspace*{-1cm}
\begin{center}
\mbox{
\epsfig{file=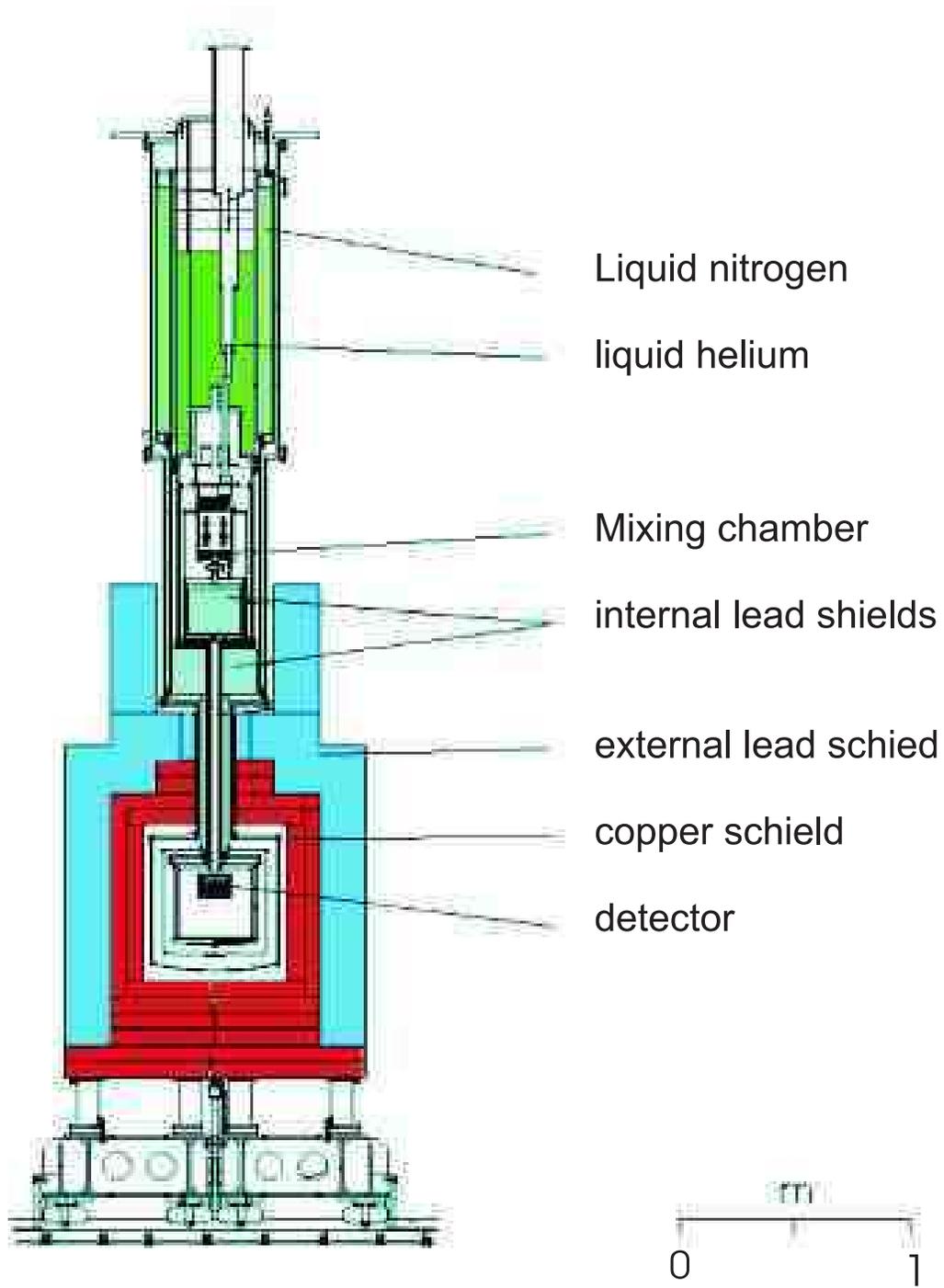,height=19cm}
}
\end{center}
\caption[]{Layout of dilution refrigerator and cold box.}
\label{fig:cryostat}
\end{figure}

Since our detectors operate at $\sim 15$\,mK, 
the central part of the CRESST low-background facility at the LNGS
is the cryostat. 
The design of this cryostat had to combine the
requirements of low
temperatures with those of low background.
The first-generation cryostats in this field were conventional
dilution refrigerators where some of  the
materials were screened for radioactivity. However,
due to cryogenic requirements some non-radiopure materials, for
example stainless steel,  cannot be completely avoided.
Therefore we chose the design shown in Fig.~\ref{fig:cryostat},
in which a well separated ``cold box'' houses the
experimental volume at
some distance from the dilution refrigerator.
The experimental volume can house
up to 100\,kg of target mass.
The cold box is made of low-background copper, with high-purity
lead used for the vacuum seals.
It is surrounded by  shielding  consisting of 14\,cm of low-background copper
and 20\,cm of lead. Special consideration was given to the space 
between the dilution refrigerator and the cold box.  
The separation was chosen large enough so that the ``neck'' of the
external shielding, together with the internal shields,
eliminates any direct line of sight from the outside world 
into the cold box. The low temperature of the dilution refrigerator
is transferred to
the cold box by a 1.5 meter long cold finger protected by thermal
radiation shields, all  of low-background copper.
A 20\,cm thick lead shield inside a copper can is placed 
between the mixing chamber and the cold finger,
with the low temperature transmitted here by the copper can.
This internal shield, combined with another one surrounding the 
cold finger, serves to block any line of sight for radiation 
coming from the dilution refrigerator into the experimental volume.

To avoid activation of the copper by cosmic rays we minimized the
amount of time 
that the copper of the shielding and the cold box  spent above
ground. After electrolytic production
the  copper was  stored in the cellar of a beer brewery near
Munich, 
shielded from cosmic rays by more than 10\,m water
equivalent.
This reduces the hadronic component of the cosmic rays by a factor
of 
about 500. Each piece was only brought out of the
brewery cellar for the few days needed for its machining,
and then returned to the cellar. 
The total above-ground exposure of the copper was about 10 weeks.

\begin{figure}[tbp]
\begin{center}
\mbox{
\epsfig{file=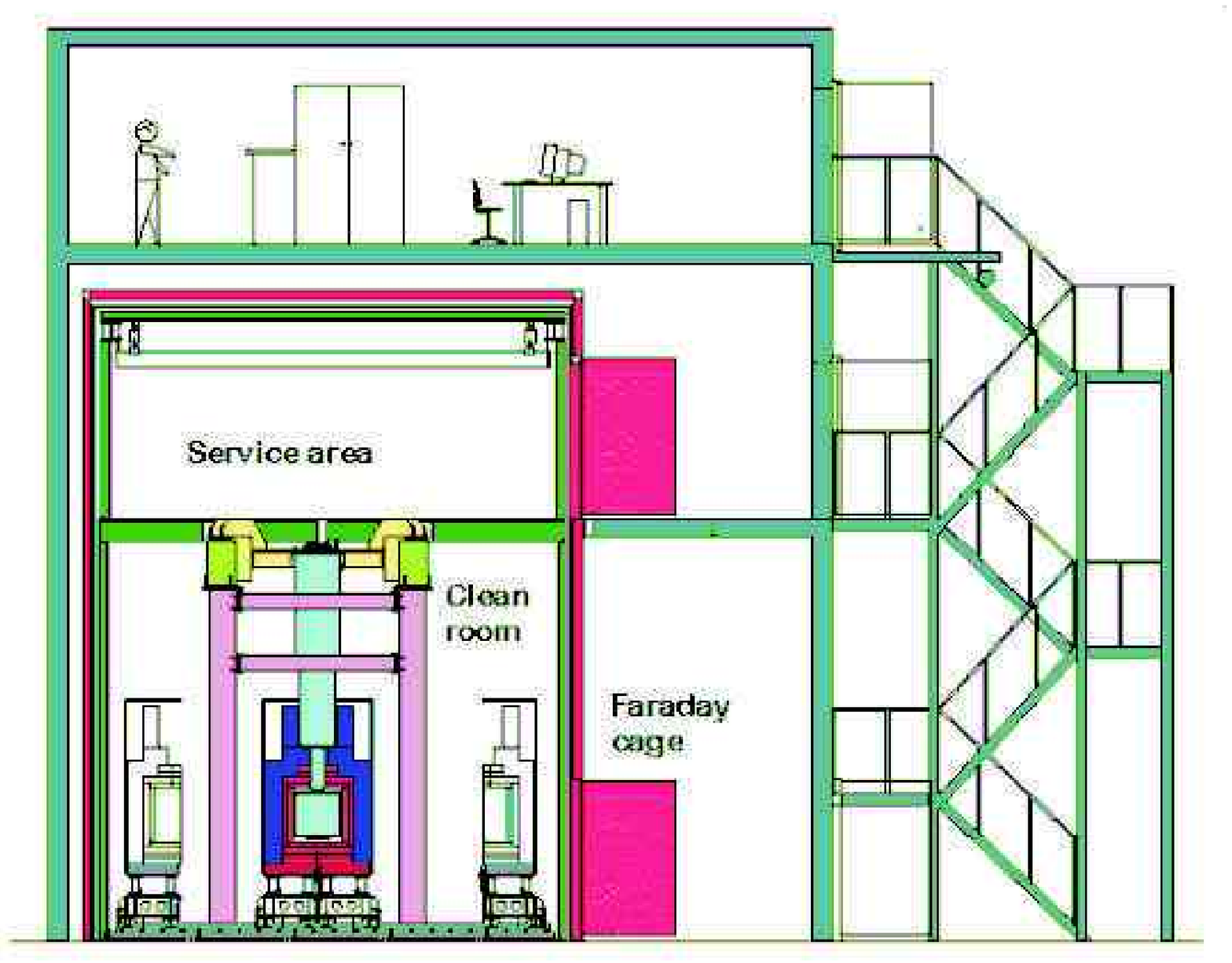,width=16cm}
}
\end{center}
\caption[]{
Cross section of CRESST building in Hall B.
The external shielding is shown in both its open and closed positions.
}
\label{FIG_hut}
\end{figure}

It is not sufficient to use high-purity materials. Their surfaces
must also be kept clean during use, 
and we have taken care to design our facilities in Gran Sasso to
make this possible.  
The Faraday cage which surrounds the experiment was chosen large
enough 
so that all work on the low-background components of the experiment
can
be performed inside the cage.  The cage is divided into two levels.
The lower level 
is equipped as a clean room with a measured clean room class of 100
to protect the 
low-background components. The external lead and copper shields are
in 
two closely fitting halves, each supported on a ``wagon'' on rails,
so that the shielding can be opened without 
handling the individual pieces.  The entire shielding is enclosed
in a gas-tight Radon box that is
flushed with N$_2$ and maintained at a small overpressure.
In its retracted position (shown in Fig.~\ref{FIG_hut})
the shielding is outside the dilution refrigerator support
structure
but still inside the clean room and sufficient room is then 
available to disassemble the cold box.

Entrance to the clean room is through a changing room external to
the 
Faraday cage (not shown in Fig.~\ref{FIG_hut}).  
 The upper level of the Faraday cage is outside the clean room and
allows access to the top 
of the cryostat for servicing and to the electronics.
To save on floor space in Gran Sasso, the counting room and a
laminar-flow work space 
for handling the detectors is placed on top of the Faraday cage.
All of this equipment is inside a building in Hall B.

The original installation used a prototype cold box,
not made of radiopure materials.
The purpose of the prototype was to test the cryogenic
functioning of the design and to provide a well-shielded
environment for completing the development of the 262\,g detectors. 
At the end of 1998
the prototype cold box was replaced  by a radiopure
version of the same design.  
After machining, the new cold box  was  cleaned by electropolishing 
and subsequent rinsing with high-purity water. The pieces were then
brought to 
Gran Sasso in gas-tight transport containers made of PE and flushed
with nitrogen. 

\section{Detectors and results of Phase I}

\subsection{Sapphire cryogenic calorimeters}

The detectors we have 
developed~\cite{Seidel,Frank,Ferger,Colling} 
consist of a dielectric crystal 
in which the particle interaction takes place,
and a small superconducting film evaporated onto the surface, 
serving as a thermometer. 
The detector is operated within the superconducting-to-normal 
transition of the thermometer, 
where a small temperature rise $\Delta T$ of the thermometer 
leads to a relatively large rise $\Delta R$ of its resistance.
The $\Delta T$ induced by a particle 
in the energy range of interest for dark matter 
is much smaller than the width of the transition, 
so that there is an approximately linear
relation between $\Delta T$  and $\Delta R$. 

We have found~\cite{model} that the energy deposited by the particle
does not thermalize in the sapphire crystal.
Instead, to good approximation, 
the high frequency phonons created by an event 
spread throughout the crystal and reflect at the surfaces
until they are directly absorbed in the superconducting film.
Thus the energy resolution is only moderately dependent 
on the size of the crystal, 
and scaling up to large detectors is feasible.

The technique can be applied to a variety of materials.
The detectors employed in Phase I of the CRESST experiment
in Gran Sasso used 
262\,g sapphire (Al$_2$O$_3$) absorbers
and tungsten (W) thermometers operating near 15\,mK. 
The 262\,g sapphire detectors were developed  
by scaling up a 32\,g sapphire detector~\cite{Colling}.  
Due to optimized design, 
and because of the non-thermalization of the phonons, 
this scaling-up was achieved 
without significant loss in sensitivity.

\begin{figure}[tbp]
\vspace*{.5cm}
\begin{center}
\mbox{\epsfig{file=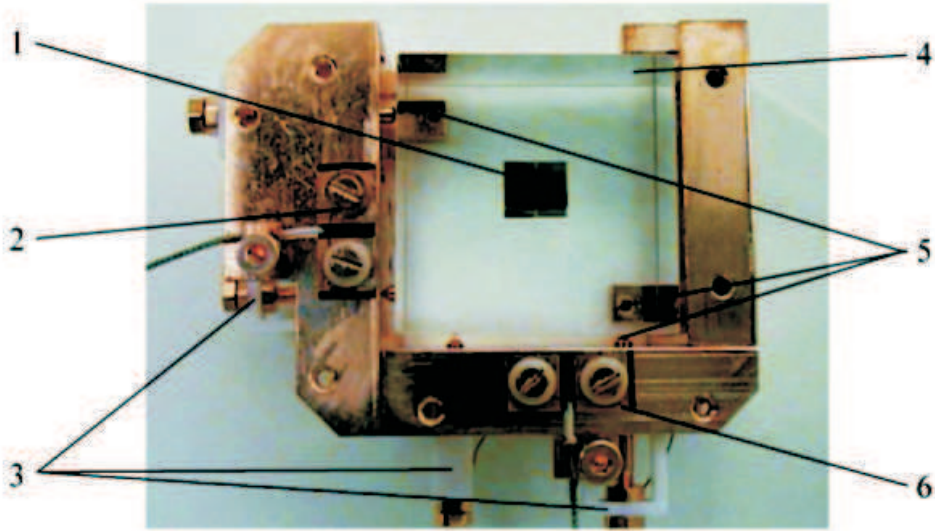, height=7.6cm, angle=0}}
\end{center}
\caption[]{
Photograph of a 262\,g sapphire detector. 
The transparent sapphire crystal (4) is in the center.
The other visible components are:
(1) tungsten thermometer,
(2) holder pads with screw contacts for connecting to the heater circuit,
(3) plastic springs,
(5) sapphire balls,
(6) holder pads with screw contacts for connecting to the SQUID 
read-out circuit. 
}
\label{fig:sapdet}
\end{figure}

Fig.~\ref{fig:sapdet} shows a 262\,g sapphire detector 
mounted in its copper holder. 
The $4 \times 4 \times 4.1$\,cm$^3$ crystal rested
thermally insulated on supports attached to the holder.
In the original design, these supports were sapphire balls.  
Some of the supports were fixed and others were 
on pins loaded with plastic springs. 

The detector used for dark matter limits had a W thermometer 
of size 3\,mm $\times$ 5\,mm. 
The electrical and thermal connections to the detector
are shown in Fig.~\ref{fig-connect}.
Thermal contact between the holder and the detector 
was provided by gold wires of diameter 25\,$\mu$m 
bonded to the Cu holder
and to a gold contact pad in the middle of the W thermometer.
The copper holder was thermally connected via the cold finger
to the mixing chamber of the dilution refrigerator,
which was stabilized to a temperature of 6\,mK.
The electrical connection to the detector 
was made by superconducting Al wires 
bonded to Al pads on each end of the thermometer 
and to isolated contact pads on the holder.  
To avoid radioactive solder joints, 
the superconducting wires from there to the 
external readout circuit were screwed to the contact pads.
The resistance of the thermometer ($\sim$0.1~$\Omega$) was measured by
passing a constant current $I_0$ through the readout circuit in which 
the thermometer was in parallel with a small ($\sim$0.05 $\Omega$) resistor
and the input coil of a dc-SQUID (Fig.~\ref{fig-circuit}).
A rise in the thermometer resistance was then measured via 
the current rise through the SQUID input coil.  

\begin{figure}[tbp]
\begin{center}
 \unitlength=1cm
\begin{picture}(10,7.2)(-5,-3.6)
\thicklines
\put(-.5,1){\line(1,0){1}}
\put(-.5,.8){\line(1,0){1}}
\put(-.5,.1){\line(1,0){1}}
\put(-.5,-.1){\line(1,0){1}}
\put(-.5,-.8){\line(1,0){1}}
\put(-.5,-1){\line(1,0){1}}
\put(-.5,1){\line(0,-1){2}}
\put(.5,1){\line(0,-1){2}}
\put(-1.05,-0.05){\framebox(.1,.1){}}
\put(.95,-0.05){\framebox(.1,.1){}}
\put(-.5,0){\oval(1,1)[t]}
\put(.5,0){\oval(1,1)[t]}
\put(1,0){\vector(1,0){1.5}}
\put(2.7,0.1){external}
\put(2.7,-0.3){heater circuit}
\put(-1,0){\vector(-1,0){1.5}}
\put(-5,0.1){external}
\put(-5,-0.3){heater circuit}
\put(.2,0){\vector(1,-1){1.5}}
\put(2,-2){thermal connection}
\put(2,-2.4){to holder}
\put(0,.9){\vector(0,1){1.5}}
\put(-.6,3.4){external}
\put(-.6,3.0){readout}
\put(-.6,2.6){circuit}
\put(0,-.9){\vector(0,-1){1.5}}
\put(-.6,-2.8){external}
\put(-.6,-3.2){readout}
\put(-.6,-3.6){circuit}
\end{picture}
\end{center}
\caption[]{Thermal and electrical connections to thermometer.
}
\label{fig-connect}
\vspace*{2cm}
\begin{center}
\unitlength=1cm
\begin{picture}(10,9)(-5,-4.5)
\thicklines
\put(0,4.5){\vector(0,-1){1.1}}
\put(0,3.5){\line(0,-1){1}}
\put(0.2,3.4){\large $I_0$}
\put(-1.5,2.5){\line(1,0){3}}
\put(-1.5,-2.5){\line(1,0){3}}
\put(0,-2.5){\vector(0,-1){1.1}}
\put(0,-3.5){\line(0,-1){1}}
\put(0.2,-3.7){\large $I_0$}
\put(-1.5,2.5){\line(0,-1){1.5}}
  \put(-2,1){\line(1,0){1}}
  \put(-2,.8){\line(1,0){1}}
  \put(-2,.1){\line(1,0){1}}
  \put(-2,-.1){\line(1,0){1}}
  \put(-2,-.8){\line(1,0){1}}
  \put(-2,-1){\line(1,0){1}}
  \put(-2,1){\line(0,-1){2}}
  \put(-1,1){\line(0,-1){2}}
  \put(-5,-.1){thermometer}
\put(-1.5,-1){\line(0,-1){1.5}}
\put(1.5,2.5){\line(0,-1){1.1}}
  \put(1.35,0.5){ \begin{picture}(1,2)(0,0)
                  \put(0,.25){\oval(.5,.5)[r]}
                  \put(0,.35){\oval(.3,.3)[l]}
                  \put(0,.45){\oval(.5,.5)[r]}
                  \put(0,.55){\oval(.3,.3)[l]}
                  \put(0,.65){\oval(.5,.5)[r]}
                  \end{picture}
  }
  \put(2.2,1.1){SQUID}
  \put(2.2,0.6){input coil}
\put(1.5,0.5){\line(0,-1){1}}
  \put(1.25,-.5){\line(1,0){.5}}
  \put(1.25,-1.5){\line(1,0){.5}}
  \put(1.25,-.5){\line(0,-1){1}}
  \put(1.75,-.5){\line(0,-1){1}}
  \put(2.2,-.8){reference}
  \put(2.2,-1.3){resistor}
\put(1.5,-1.5){\line(0,-1){1}}
\end{picture}
\end{center}
\caption[]{
Readout circuit to measure the resistance of the thermometer.
}
\label{fig-circuit}
\end{figure}

In a separate circuit, 
a heater to control the temperature of the detector 
was provided by a $\sim 5$\,mm long 25\,$\mu$m diameter gold wire
which was bonded to
the gold pad in the center of the W thermometer  
and two very small Al contact pads on the sapphire crystal 
to either side of the thermometer.
External connections to the two small Al pads were used to apply a 
controlled voltage across this gold wire.
To avoid interaction between the heater circuit 
and the readout circuit, 
the place where they connect -- the bond spot of the gold wire --
was made as small as possible
and its long axis was perpendicular to 
the direction of current flow in the thermometer.
The thermometer temperature was kept constant between pulses
using the baseline of the SQUID output voltage as the temperature indicator
and regulating the voltage to the heater 
under computer control using a proportional integral algorithm. 
The heater was additionally used to inject short heat pulses 
for monitoring the long term stability of the energy calibration 
and for measuring the trigger efficiency close to threshold.

\begin{figure}[tbp]
\begin{center}
\mbox{  
\epsfig{file=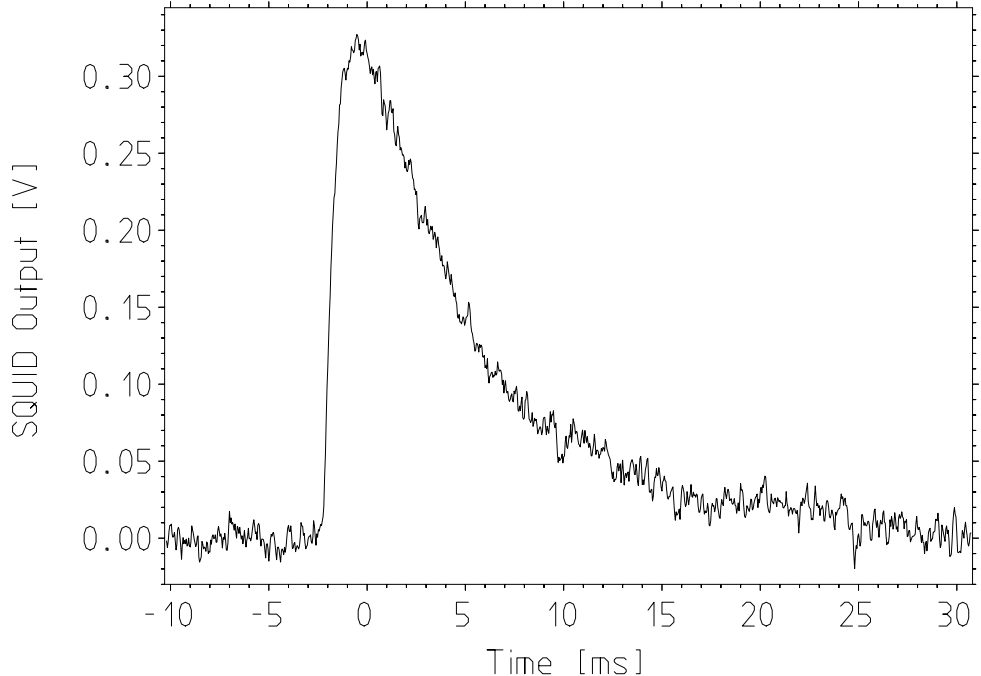,height=13cm,angle=90}
}
\end{center}
\caption[]{Typical measured pulse of about 6\,keV.
}
\label{fig-pulse}
\end{figure}

For the data-acquisition system, 
the output voltage of the SQUID electronics was split into two branches.
One was shaped and ac-coupled to a trigger unit and the other passed
through an anti-aliasing filter and 
was dc-coupled to a 16-bit transient recorder.
The time base of the transient recorder was chosen to 
be $40\,\mu$s, which provided about 20 samples in the rise time
of the pulse.
The record length of 1024 time bins included
a ``pre-trigger'' region of 256 bins, 
to record the baseline before the event, 
and a ``post-trigger'' region which contained the pulse
(Fig.~\ref{fig-pulse}).
The transient recorder data 
for each triggered event were written to disk for off-line analysis.
After each trigger there was a dead time of $\sim 25$\,ms to allow time 
for the readout and the next pre-trigger region.
Pulses arriving in another detector within half of the post-trigger period 
of the detector which triggered first were also recorded, 
including the time delay with respect to the first trigger.

\subsection{Data taken in Gran Sasso}

The sensitivity and size of the 262\,g detectors 
meant that they could only be meaningfully tested 
in a low-background environment.
This was first done in our setup in Gran Sasso 
using the prototype cold box of normal copper.
Using active thermal feedback,
an energy resolution of 133\,eV (FWHM) at 1.5\,keV for X-rays
was achieved~\cite{Meier}.
This active feedback was not used in our dark matter run,
and without it the resolution at 1.5\,keV is more typically 
230-330\,eV~\cite{SistiNIM}.

During 1999, a first series of measurements with four 262\,g
detectors under low-background conditions was performed 
in the new radiopure cold box.  
The measured background was much higher than expected.
It was time-dependent and not Poissonian,
indicating that it was not caused by radioactivity.
The origin of this background was investigated 
in a series of runs 
and finally identified as
the spontaneous formation of microscopic cracks in the sapphire crystal 
at the points where it was supported by sapphire balls. 
Due to the extremely small contact area of the balls,
an excessive pressure resulted from the force 
needed to tightly hold the crystal. 
In the spring of 2000 the balls were replaced 
by plastic stubs with a larger contact area (3 mm diameter) 
and the spurious background completely disappeared. 
The use of these stubs did not lead to a noticeable 
loss of sensitivity, despite their larger contact area.  

To study the background and obtain dark matter limits, 
several runs were performed in 2000, 
with the longest one lasting about 3 months.
The high reliability, long-term stability and up-time during these runs 
demonstrated convincingly the suitability of such a system for
dark matter searches.

In October 2000 the shaping of the trigger signal was optimized 
and a lower trigger threshold was obtained.
A week of data were taken under these conditions.
Due to this lower threshold, the dark matter limits obtained from these
data are better than those from the previous longer runs,
and it is these data which we discuss further.
The data consist of a 10-hour calibration run 
with an external $^{57}$Co source,
138.8 hours of data without source and finally  another calibration run.
The 138 hour run will be used to set our dark matter limits. 

\subsection{Detector monitoring and calibration}

The performance of the detector was monitored by heater pulses
injected into the small heater wire bonded to the W thermometer.
These are produced by a voltage pulse from a pulser module,
with the shape adjusted to create a detector response 
similar to that caused by a particle interaction. 
A pulse was sent every 30\,s throughout both dark matter and calibration runs.
The height of the pulses was varied to cover the whole dynamic range, 
with more pulses in the low energy region.
This method provides a monitor of the stability of the detectors,
an extrapolation of the energy calibration over the whole dynamic range,
and a measure of their trigger efficiency as a function of deposited energy.

\begin{figure}[tbp]
\begin{center}
\mbox{  
\epsfig{file=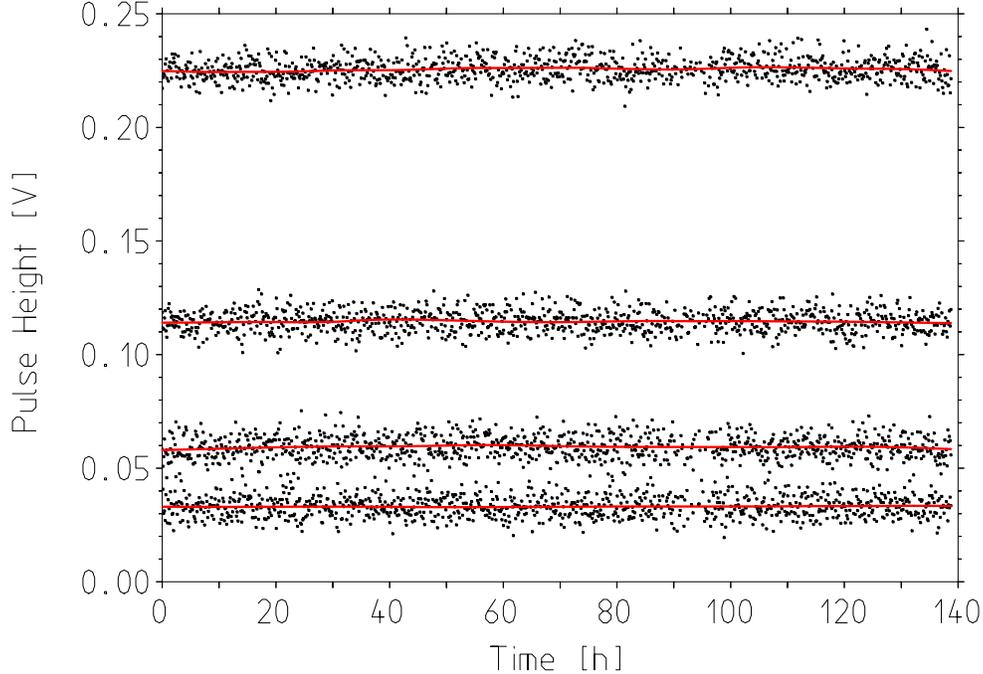, height=13cm, angle=90}
}
\end{center}
\caption[]{
The measured pulse height of detector \#\,8 
as a function of time during the dark matter run
for the heater pulses of energy 0.58, 1.04, 2.04, and 4.08\,keV.
The detector is seen to be stable to within the resolution.
The fitted lines are used to calculate the response function at
the time of event pulses as shown in 
Fig.~\ref{fig-calib}.   
}
\label{fig-stability}
\end{figure}

The individual detectors varied in their response, 
with detector \#\,8 
(numbered by order of fabrication)
having a lower threshold 
and thus giving the best dark matter limits.
The trigger efficiency of this detector was measured to be 100\,\% 
down to an energy of 580\,eV throughout the data 
used to extract the dark matter limits.
The stability of detector \#\,8 during the dark matter run
can be seen in Fig.~\ref{fig-stability}, where the response 
to heater pulses is shown. 

\begin{figure}[tbp]
\begin{center}
\mbox{  
\epsfig{file=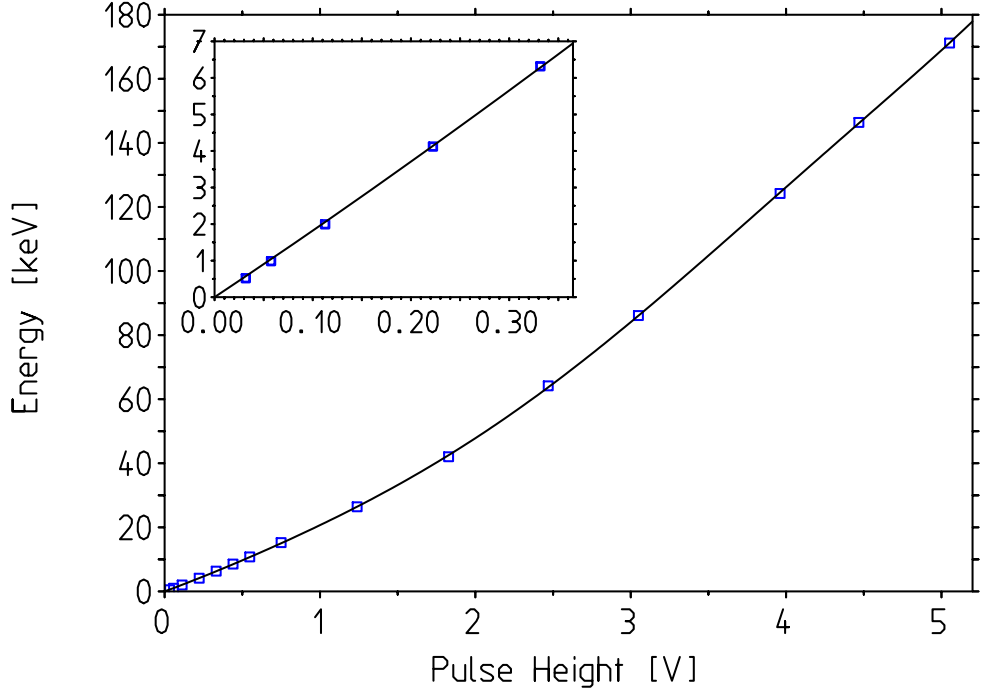,height=13cm,angle=90}
}
\end{center}
\caption[]{The pulse heights for detector \#\,8 for each injected 
heater pulse energy calculated from the lines in 
Fig.~\ref{fig-stability}  
are shown as points. The curve is the fitted polynomial which is used as 
the response function. The insert is an enlargement of  the low energy
region.
}
\label{fig-calib}
\end{figure}

To calibrate the energy scale 
a $^{57}$Co source (122 and 136 keV $\gamma$ lines) 
was inserted inside the shielding via a removable plug,
illuminating the cold box from below.
Data were taken with this source along with the heater pulses.
Comparison of the pulse heights from the source and heater pulses
provides an absolute calibration of the heater pulses in terms of 
equivalent $\gamma$ energy.
The amplitude of each pulse was determined by fitting it to a template.
This avoids the bias of picking the highest point of suitably filtered pulses,
which is systematically pulled by noise fluctuations to larger values. The absence
of any  bias is important for a precise definition of the threshold. It   
was therefore checked by fitting empty base lines randomly sampled 
during the whole data taking period. The  resulting distribution 
peaked without any bias at (-0.0019 $\pm$ 0.003)\,keV.

The templates were made by averaging many pulses from the 
same heater pulse voltage. 
A separate template was made for each heater energy, 
and for source pulses around the Compton edge (30-35\,keV).
To extrapolate the energy calibration over the whole dynamic range,
we used the heater pulses and plotted their fitted amplitude versus
the injected energy (Fig.~\ref{fig-calib}).
These data were fit with a polynomial function 
to give the detector response as a function of deposited energy.

For the dark matter data,
the response function determined above was used
to convert each recorded event pulse 
to energy in each time bin.
An optimal filter was then used on this converted pulse 
to determine the pulse height.
This gave a slightly better resolution than a template fit.
A comparison to the template fit showed that the optimal filter 
applied to the linearized pulses does not introduce an energy bias.
The resulting spectrum for the dark matter run is shown as the 
upper histogram in Fig.~\ref{fig-cuts}.

The reliability of the energy calibration method 
to low energy was checked
with a dedicated run where a low activity $^{57}$Co source was 
mounted inside the cryostat directly facing the crystals. Besides the
122 and 136 keV $\gamma$ emission lines, 
this source gave a 14.4\,keV $\gamma$ line and a 6.4\,keV Fe X ray line. 
The source was chosen to be very weak to reduce the chance 
of contamination so that one week run gave only a small number of counts
in the 14.4 and 6.4\,keV lines.
After applying the standard calibration method of extrapolation 
from the 122\,keV line as described above, 
the measured energies for the 14.4 and 6.4\,keV
lines were  15.16$^{+0.09}_{-0.09}$ keV and   6.70$^{+0.07}_{-0.05}$ keV, respectively,
with the fit errors corresponding to 90 \% CL. 
Our calibration procedure puts the 14.4 and the 6.4\,keV lines  5.3 \% and 5.4\% too high.
Since it is the lower energies which most affect our dark matter limits,
this tendency to shift events up in energy will put our limits on the 
conservative side.

\subsection{Energy spectrum from dark matter run}

The data used to set dark matter limits
were taken during a week in October 2000,
with a few short interruptions to re-fill the cryostat with 
liquid helium.  The total run time was 138.8 hours, 
of which 0.6 hours is dead time following triggers.

To avoid reliance on  the detailed behaviour of the trigger efficiency 
at very low energies a software threshold of 600\,eV was used.
There were  374 events from the software threshold to 20\,keV.

Events in coincidence in two or more detectors
cannot be due to WIMP interactions, and
so can be discarded.
The time difference  distribution between events 
showed a clear coincidence peak with a width of about 0.4~ms and long tails
extending to about 2 ms for very low detected energies.
The coincidence cut was set at $\pm$ 4 ms, removing 73 events.
Considering that only two detectors were active in this run and  
only one out of six faces was facing the other crystal, the
coincidence rate of almost 20\% is surprisingly high.

\begin{figure}[tbp]
\begin{center}
\mbox{  
\epsfig{file=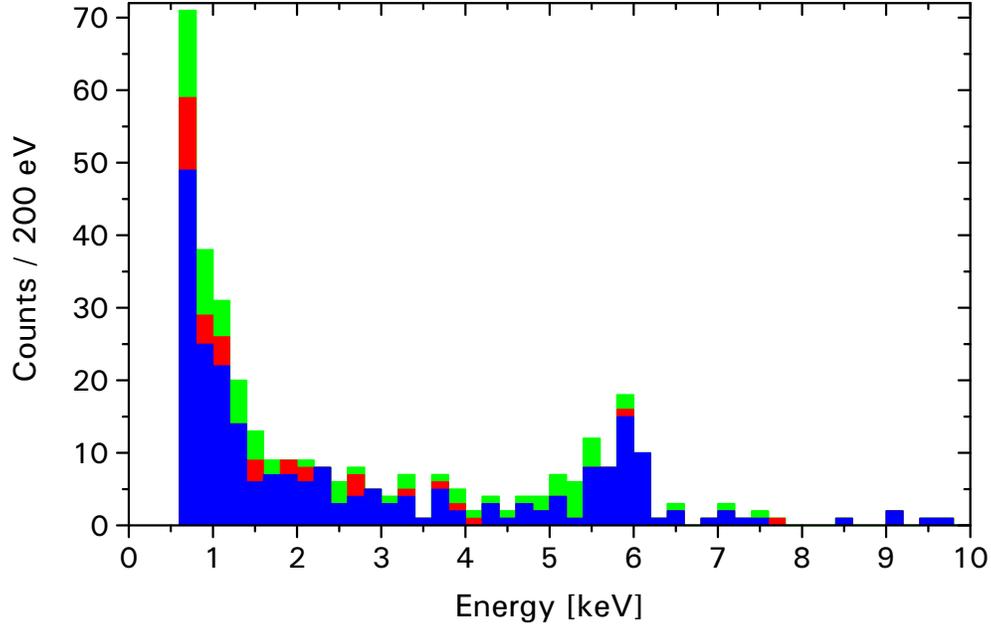, height=13cm,angle=-90}
}
\end{center}
\caption[]{
Energy spectrum of events in the dark matter run
(without source) in 200\,eV bins.  
The upper histogram shows the uncut data,
the middle histogram the data after coincident events are rejected,
and the lower histogram after the pulse shape  cut.
}
\label{fig-cuts}
\end{figure}

The pulse shape of the remaining 301 events was then examined.
Some of the events were spurious, induced by 
mechanical vibration or electronic noise,
showing an abnormal pulse shape.
Particle interactions can also produce distorted shapes 
when the energy deposit is high and beyond the detector's dynamic range. 
To judge the correctness of the pulse shape, each event 
was fitted to a template and the r.m.s. deviation calculated.
A cut on this deviation was chosen to be conservative and have a 
retention efficiency
of essentially 100\,\% at all energies for good events.
Its efficiency was tested with heater pulses, 
resulting in only 1 out of 1032 of the 580\,eV heater pulses being  
discarded.
A second test with the calibration source gave
that only 0.22\,\% of the events were discarded.
After the pulse shape cut 265 events remain.
These are shown as the lower histogram in Fig.~\ref{fig-cuts}.

\begin{figure}[tbp]
\begin{center}
\mbox{
\epsfig{file=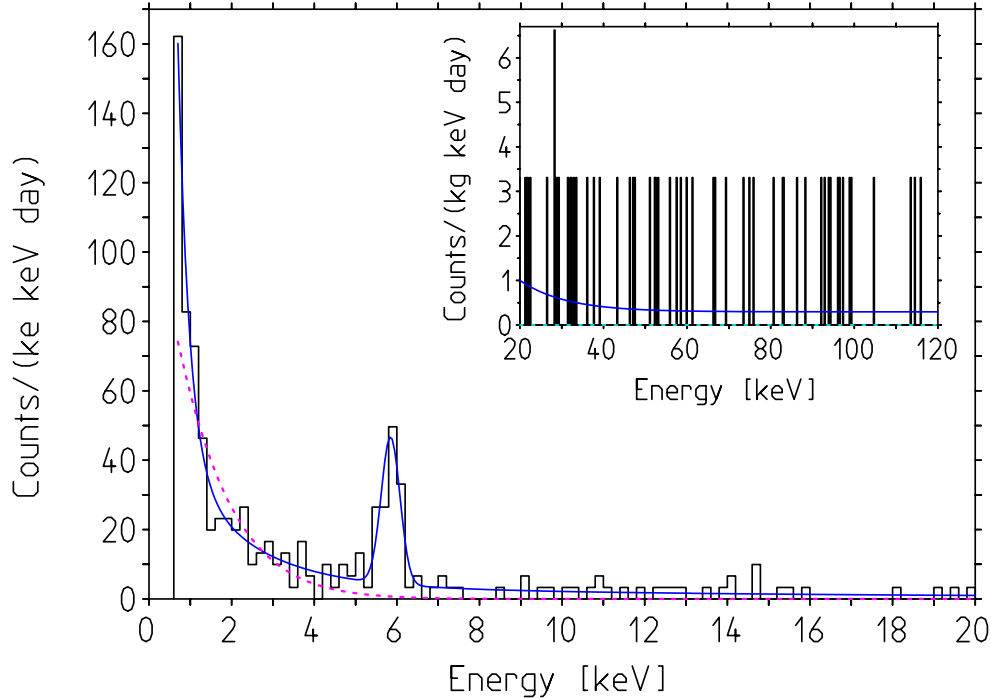,height=13cm,angle=90}
}
\end{center}
\caption[]{
Energy spectrum of detector \#\,8 during the dark matter run 
(1.51 kg days) in 200\,eV bins.The insert shows the spectrum at higher energies.
The fully drawn curve is an empirical fit to the experimental spectrum 
which serves for extracting dark matter limits.
For illustration a 5\,GeV WIMP excluded at 90\,\% C.L is shown as a dashed curve.
}
\label{fig:cresst1_spectrum}
\end{figure}

The final spectrum for detector \#\,8 is shown again 
in Fig.~\ref{fig:cresst1_spectrum}.
In the energy range from 15 to 25 keV  the background is 
$(0.73\pm 0.22)$ counts/kg/keV/day and  drops
to about 0.3 counts/kg/keV/day at 100\,keV.  

The spectrum shows a peak at about 5.9\,keV with 
$(7.0 \pm 1.2)$ counts/day. 
The position of the peak suggests a contamination with $^{55}$Fe
in the vicinity of the crystal.
$^{55}$Fe emits Mn X-rays at 5.9\,keV with no $\gamma$'s
and was indeed used as a source 
for characterizing the performance of the detectors.
The spectra measured with other detectors are very similar 
with nearly the same count rates in both the peak and the 
continuous part of the spectrum. 

\subsection{Limits on WIMP dark matter}

The analysis for dark matter limits of the two possible WIMP
interaction cross sections for our sapphire detectors, i.e. 
spin--independent $\sigma_{scalar}^{W-N}$ and spin-dependent
$\sigma_{axial}^{W-N}$ WIMP--nucleon cross sections, searches for the
cross section which is just large enough that the amount of  WIMP events 
becomes inconsistent with the measured spectrum at a given confidence level 
(here 90\% CL). No general algorithm for such an analysis has been
found in the literature. Therefore we developed two conceptually
independent approaches to be described below and checked for
consistency of the results. 
\par
For calculating the energy spectrum of nuclear recoils from elastic collisions 
between WIMPs and the nuclei of the detector we use formulas from
the extensive reviews \cite{lewin,jungman} for a truncated Maxwell
velocity distribution in an isothermal WIMP--halo model. The parameters 
which have been used are summarized in table~\ref{parameters}. 
For the spin dependent interaction channel only the  $^{27}$Al nuclei  with a 
spin of 5/2 and100\,\% natural isotipic abundance contribute, while for spin independent 
interaction Al and O nuclei both contribute with a rate proportional to $A^2$.
\begin{table}[htb]
\begin{center}
\caption{List of parameters used for calculating WIMP spectra}
\vspace*{3mm}
\begin{tabular}{ll}
Parameter name & value \\\hline
WIMP velocity distribution & 270 km/s \\
Escape velocity & 650 km/s \\
Earth relative velocity & 230 km/s \\
WIMP local halo density & 0.3 GeV/cm$^{3}$ \\
\end{tabular}
\label{parameters}
\end{center}
\end{table}
\par
The total WIMP--nucleus scattering cross section $\sigma(q)$  at  finite momentum 
transfers $q$  is parameterized as 
$\sigma(q)=\sigma{_0} F^2(q)$, where $\sigma_{0}$ is the cross section at zero
momentum transfer and  F(q) is the form factor which accounts for the loss of coherence 
at larger momentum transfers.  The Helm form 
factor \cite{Helm} with the modifications proposed in \cite{lewin} has been selected 
for both interaction channels. For small nuclei like $^{27}$Al the momentum 
transfer is small for all WIMP masses, and details of the form factor have 
negligible effect on the resulting exclusion plot. Therefore the selected form factor is 
adequate for both interaction channels. 
The energy resolution at the 5.9\,keV peak in Fig.~\ref{fig:cresst1_spectrum}.
is $\delta E$ = (572 $\pm$ 90) eV (FWHM), 
whereas the resolution of the Fe K$_{\alpha}$ peak at 6.4 keV 
from the internal calibration source is (200 $\pm$ 50) eV in agreement with the
energy resolution of the heater pulses. At the 122 keV peak of the calibration source
the resolution degrades to about 5 keV. To account for the finite energy resolution of
the detector the recoil spectrum was convolved with a Gaussian
with an energy dependent full width at half maximum of 
$\delta E = \sqrt{a^2 + b^2  E ^2}$. 
With a=0.52 keV and b=4.1, this gives $\delta E$ = 0.57\,keV and $\delta E$ =5\,keV 
at $E$=5.9 keV and $E$=122 keV, respectively. 
We further assumed a quenching factor of 1, i.e. 100\% of the nuclear recoil energy 
is detected in the phonon readout channel.  An experimental proof of this
plausible assumption does not exist up to now for our sapphire detectors. However, 
quenching factors close to 1 have been measured with other cryogenic 
detectors \cite{fiorini}. 
\par
\begin{figure}[tbp]
\begin{center}
\mbox{
\epsfig{file=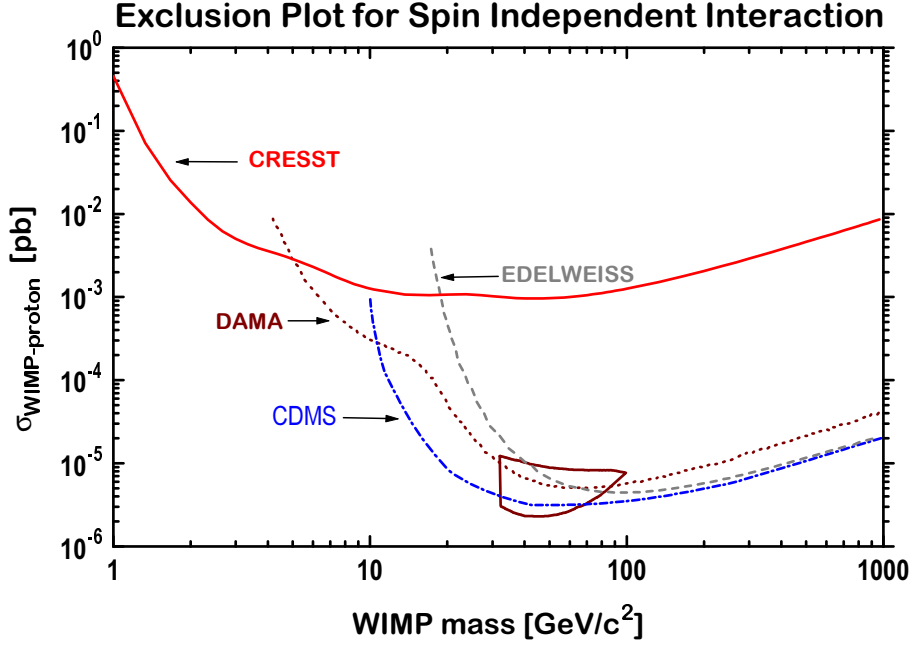,height=12cm,angle=-90}
}
\end{center}
\caption[]{
Equivalent WIMP-nucleon cross section limits (90\,\% CL) 
for  spin-independent interaction as a function of the WIMP mass 
for 1.51 kg days exposure of a 262\,g sapphire detector. 
For comparison  DAMA NaI limits with pulse shape discrimination~\cite{DAMA_sd}, 
CDMS limits with statistical subtraction of the neutron background~\cite{cdms2000},
limits from the  UK dark matter search~\cite{UK_sd} and from the EDELWEISS WIMP search
with a heat and ionization Ge Detector~\cite{EDELWEISS_2001} are also shown together with 
the allowed region al 3 $\sigma $ CL for 
a WIMP r.m.s velocity of 270 km/s from the DAMA annual modulation data ~\cite{ritapositive}.
}
\label{fig-si_exclusion}
\end{figure}
\begin{figure}[tbp]
\begin{center}
\mbox{
\epsfig{file=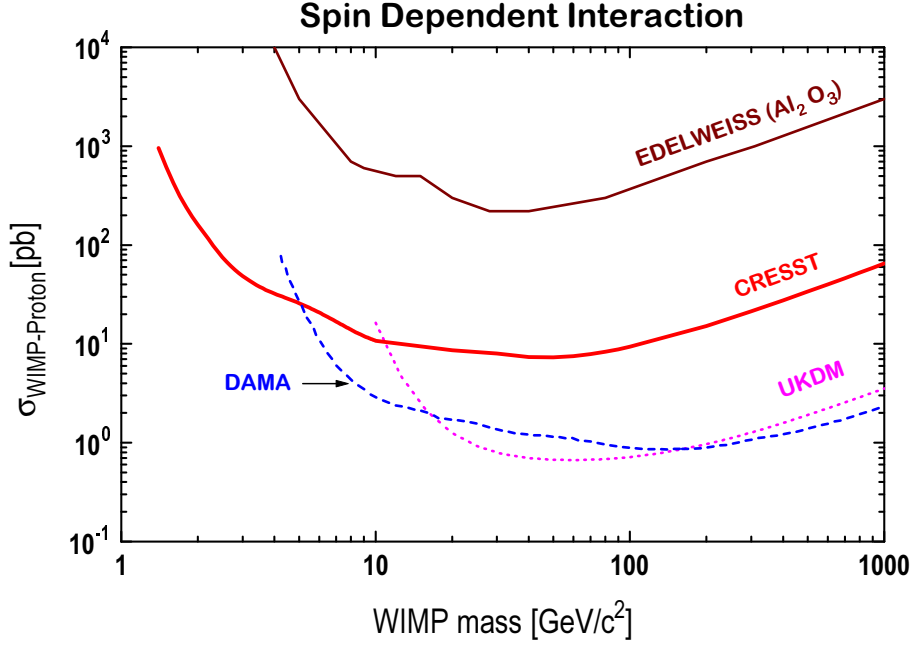,height=12cm,angle=-90}
}
\end{center}
\caption[]{
Equivalent WIMP-proton cross section limits (90\,\% CL) 
for spin-dependent interaction as a function of the WIMP mass 
from 1.51 kg days exposure of a 262\,g sapphire detector. 
For comparison we show  limits from the EDELWEISS dark matter search 
with cryogenic sapphire detectors~\cite{EDELWEISS_sap}, 
from  DAMA with NaI detectors using pulse-shape discrimination~\cite{DAMA_sd}, 
and from the UK dark matter search with NaI detectors~\cite{UK_sd}.
}
\label{fig-sd_exclusion}
\end{figure}

The analysis procedure which has been used to obtain the exclusion
curves is Fig.~\ref{fig-si_exclusion} and Fig.~\ref{fig-sd_exclusion} works as follows: 
After calculating the
shape of the energy spectrum for a given WIMP mass, $\sigma_{0}$ 
which is the scale factor for the intensity of the WIMP expectation,
 is determined by a  maximum--likelihood comparison of the calculated WIMP 
spectrum with the measured energy spectrum. The analysis is aiming to find the 
amount of WIMP events that just too large to be hidden under the measured 
spectrum. 
Some energy intervals
(typically close to the threshold) are more effective than others for
constraining the existence of a WIMP signal in the data. Therefore, it
is reasonable to select these energy intervals and use them to give
the most stringent exclusion limits on a WIMP signal that can be
obtained from the data. 
\par
We performed such a search, using a sliding,
variable-width energy window with a smallest width of 1.2\,keV. A similar method
was used in \cite{hdmos,rosebud, EDELWEISS_sap}.  
Since the method tends to  
pick a 'lucky'  downward fluctuation  in the data, especially in a low count
rate spectrum, it  may have a significant statistical bias.  
Therefore we have investigated the selection bias with two 
simple alternative methods.

\begin{figure}
\begin{center}
\mbox{\epsfig{file=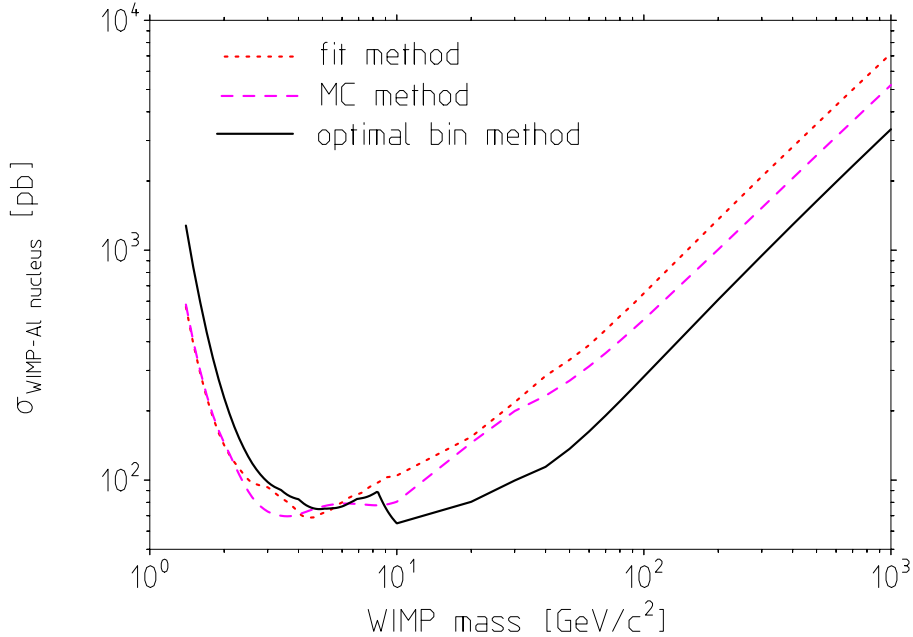,height=12cm,angle=90}}
\caption[]{Comparison of different statistical methods to extract exclusion limits. 
Shown is the  WIMP-$^{27}$Al nucleus cross section limit
 (90 \% CL)  obtained with 3 different methods. The MC-method was selected to extract the  
limits shown in Fig.~\ref{fig-sd_exclusion} and Fig.~\ref{fig-si_exclusion}.}
\label{fig-methods}
\end{center}
\end{figure}

In the first method we fit the spectrum  with an  empirical  function 
$B$ as shown in Fig.~\ref{fig:cresst1_spectrum} which results in a likelihood of 
$\mathcal{L} = \mathcal{L}_{0}$.  Without a WIMP  contribution this function 
just describes the background.
When a small WIMP signal $S$  is first added, which is everywhere smaller then the fit 
function, then with a re--definition of the background to $B` = B - S$  we can achieve to 
leave  $\mathcal{L}$ unchanged.  With increasing $\sigma_{0}$ , 
when $S$ starts to exceed the fit function in the first energy bin, this compensation  
would require a negative background and in this bin the function to compare with the 
data is then the WIMP signal and  $\mathcal{L}$ starts to decrease. 
The cross section excluded at 90\% CL is obtained when 
 $\mathcal{L} = \mathcal{L}_{0}$ - 1.28$^2$/2 is reached.
\par
The second method starts from the same empirical fit of the data and then uses this fit function
as the poissonian mean in a Monte Carlo calculation of a large number of synthetical spectra. 
These spectra are different statistical realizations of the  measurement when the fit function 
is adequate for describing the data. 
The same function is then fit to each synthetical data set to probe the statistical variation 
of the fit curve. Then, from each fitted curve $\sigma_{0}$ is determined such as to  make the WIMP 
signal just large enough to explain the fit curve in one energy bin. The $\sigma_{0}$ which is
excluded at 90\% CL is then obtained as the cross section for which 90 \% of the values 
determined from all fits are below. Finally, the adequacy of the fit function 
can be inferred from a comparison 
of the likelihood from the fit of the data with the likelihoods from the fits of 
the synthetic data sets.  For the curve shown in  Fig.~\ref{fig:cresst1_spectrum}, 49.1\% of the  
synthetic likelihood values are larger, indicating a perfect choice of the fit function.  

The results of these methods are compared in Fig.~\ref{fig-methods}. 
At low WIMP masses the optimal bin method yields weaker limits. 
This is connected with the restriction to a 1.2 keV minimal width of 
the sliding bin which is quite large for very low WIMP masses for which 
the optimal bin starts at threshold.  For WIMP masses above 10 GeV on the other hand
the optimal bin is searched in relatively wide region with low statistics
which results in a large selection bias.  For a WIMP mass from 30\,GeV to 1000\, GeV 
for example the region from  22.7\,keV to   26.3\,keV with no counts inside is selected 
as the optimal bin.
To avoid this bias we have chosen the MC method. The small difference between MC 
and fitting method still needs further investigation and therefore we consider the 
results presented here still as preliminary. 

\par
In order to compare our results to those of other experiments, we have to
normalize the obtained WIMP--nucleus cross sections (for
spin-dependent, axial, and spin-independent, scalar, interactions) to WIMP--nucleon cross
sections. Following \cite{lewin,jungman}, the normalization for the
scalar interaction channel is straight forward and yields the results shown in
Fig.~\ref{fig-si_exclusion}. The scalar channel is not very favourable for a
target with light nuclei like sapphire, containing aluminium and oxygen, 
due to the crucial $A^{2}$ coherence factor, especially for
higher WIMP masses. That will change with the CRESST-II  detectors  as explained in
the next section.

\par
For axial interactions, only $^{27}$Al has to be considered, since it
has spin $5/2$.  The spin dependent total cross section 
at  zero momentum transfer in the notation of Ref.~\cite{tovey} is given by 
\begin{equation} 
\sigma_0 = 4 G_F^2 \mu_A^2 C_A ,
\end{equation}
where the WIMP-target reduced mass $ \mu_A$ is given by  
$\mu_A = m_A m_\chi  / (m_A + m_\chi )$ for WIMP mass  $m_\chi$  and
target nucleus mass $m_A$. The spin factor $C_A$  is given by 
\begin{equation}
C_A = \frac{8}{\pi} [a_{p}\,\langle{}S_{p}\rangle{} +
a_{n}\,\langle{}S_{n}\rangle{}]^2 \frac{J+1}{J} . 
\label{eq-spin-factor}
\end{equation}
where $a_p$ and $a_n$ are (WIMP-type dependent) effective WIMP-proton und 
WIMP-neutron couplings 
and $\langle{}S_{p,n}\rangle{}$ are the expectation values of the proton and 
neutron spins within the nucleus with total nuclear spin J.
 The limit $\sigma_{0}$ obtained for the WIMP- $^{27}$Al nucleus cross section can 
be translated  into a limit for the 
WIMP-proton  cross section by the conversion
\begin{equation}
\sigma_ {WIMP-p} = \sigma_{0} \times  \frac{\mu^2_p}{\mu^2_A} \times
 \left(  \frac{C_p}{C_A} \right).
\label{eq-normalization}
\end{equation} 
%
 However, this normalization turned out to be nontrivial as the
recent discussions in the literature has shown \cite{lewin,tovey}. The
'conventional' procedure relied on the odd-group model which uniquely
determines the spin factor for a given material, see e.g. Tab.~3 in \cite{lewin}. 
Furthermore, the normalization conveniently cancelled the WIMP model 
dependencies coming from particle 
physics. That can be directly seen from the spin factor in Eq.\ref{eq-spin-factor}, 
where the $a_{p,n}$ parameters contain the particle physics model-dependencies. 
\par
For free nucleons the spin factor is \cite{tovey}
\begin{equation}
C_{p,n}=\frac{8}{ \pi} \times \frac{3}{4} \times \,a_{p,n}^{2}. 
\end{equation}
 The odd-group model implies that  for nuclei with 
unpaired protons or neutrons either $\langle{}S_{p}\rangle{}$
or $\langle{}S_{n}\rangle{}$ is zero in Eq.~\ref{eq-spin-factor}, respectively.
Therefore, when normalized to the unpaired nucleon the model dependent 
$a_{p,n}$ factors conveniently cancelled  in Eq.\ref{eq-normalization}.  According 
to  shell model calculation (see \cite{jungman,lewin,tovey} and
references therein; for $^{27}$Al, see \cite{engel})  strict p- or n-type 
nuclei do not exist and 
in general both, $\langle{}S_{p}\rangle{}$ and $\langle{}S_{n}\rangle{}$  
contribute and can even interfere in Eq.~\ref{eq-spin-factor}.

As proposed in \cite{tovey}, an experiment should therefore
publish results separately for the p-type part and the n-type part for
their particular nucleus. However, in order to be able to compare our
results to already published limits, we applied the odd-group model
spin factor, although the value in \cite{lewin} for aluminium is
outdated. The more recent spin factor using the numbers from
\cite{engel} would result in a factor 1.9 reduction of the whole
exclusion curve to lower cross sections, in case one neglects the
neutron part for the spin factor as described in \cite{tovey}. 
One should be aware,
however, that using the odd group model approach any comparison 
to n-type nuclei like $^{73}$Ge becomes very problematic. Such a comparison 
inevitably involves model-dependencies (see e.g. the discussion in \cite{lewin} and their
Tab.~4).  All limits in Fig.~\ref{fig-sd_exclusion}  consequently come from p-type nuclei
(aluminium, sodium and iodine). As can be seen in Fig.~\ref{fig-sd_exclusion}, we improve
existing limits in the low-mass region. 

\section{Plans for Phase II}

\subsection{Scintillating calorimeters}

Passive techniques of background reduction --  
the deep underground site, 
efficient shielding against radioactivity of the surrounding rocks,
and use of radiopure materials -- 
cannot completely eliminate background from 
radioactive contaminants inside the detectors and their surroundings. 
$\beta$ and $\gamma$ absorption produce electron recoils, 
whereas WIMP and neutron scattering produce nuclear recoils.
Therefore a significant improvement in sensitivity 
can be achieved if, 
in addition to the usual passive shielding, 
the detector can distinguish electron recoils from
nuclear recoils and reject them.

It is well
known that at room temperatures nuclear recoils in  scintillators
produce much less
scintillation light than electron recoils.
Thus the combination
of a scintillator, measuring light,
and a cryogenic detector,  essentially measuring 
 total energy, can  discriminate between nuclear and electron
recoils by using  the ratio of detected light  to thermal  energy.
We have investigated the light output  of several scintillators at
mK temperatures, concentrating on inorganic intrinsic scintillators
since their scintillation efficiency usually increases at  lower
temperatures.
All scintillators so far tested  (BGO, BaF$_2$, PbWO$_4$, CaWO$_4$)
appear to function adequately at mK temperatures. 

For an initial test~\cite{meunier}
of a scintillating calorimeter we chose CaWO$_4$.
The test setup consisted of two independent detectors. 
Each one was of the CRESST type: a tungsten
superconducting phase transition thermometer with SQUID readout.
The scintillating absorber and phonon detector was a 6\,g CaWO$_4$ crystal.
The light detector was a sapphire wafer,
coated with silicon on one side to enhance light absorption.  
The CaWO$_4$ crystal 
was irradiated with the 122\,keV and 136\,keV photons from a
$^{57}$Co-Source 
and simultaneously with the $\beta$ spectrum from a $^{90}$Sr source,
with the two sources contributing about equally to the count rate. 
In addition an americium-beryllium source provided neutrons.
The trigger was given by the phonon detector.

\begin{figure}[tbp]
\begin{center}
\mbox{\hspace*{-4cm}
\epsfig{file=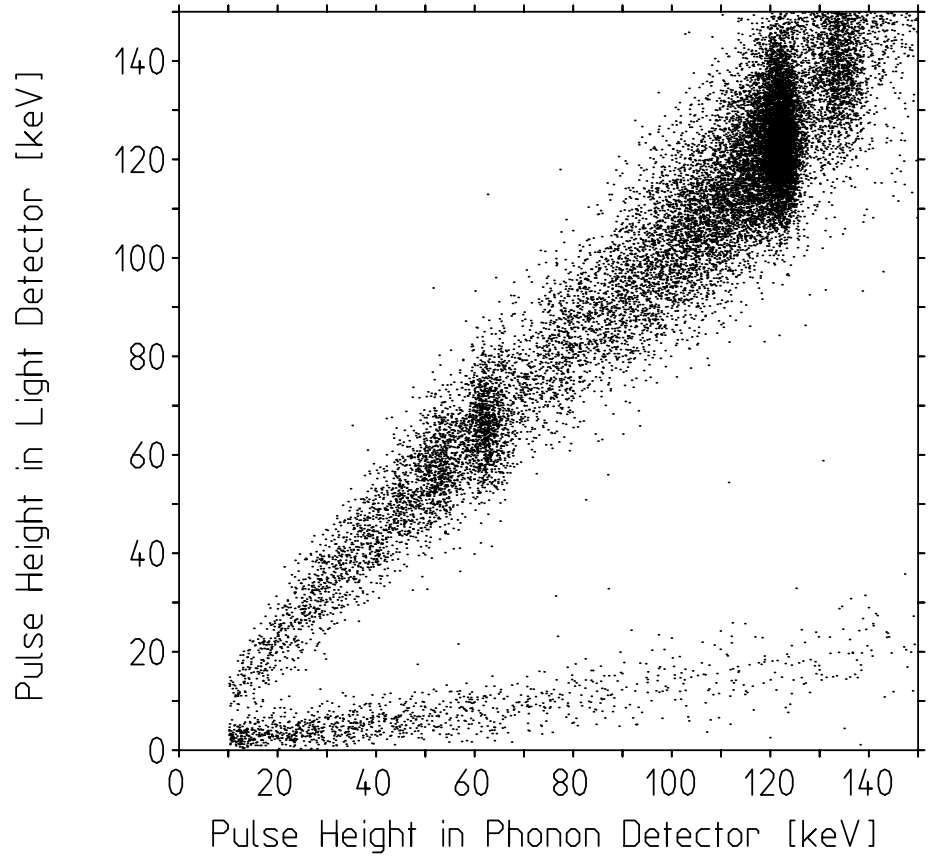,height=16cm,angle=90} 
}
\end{center}
\caption{Pulse height in the light detector versus pulse height in
the 6\,g CaWO$_4$ phonon detector, 
measured while the detector was irradiated with
photons, electrons, and neutrons.  
The lower band is caused by neutron-induced nuclear recoils,
the diagonal band by recoils from photons and electrons.
}
\label{fig:gamcorr} 
\end{figure}

Fig.~\ref{fig:gamcorr} shows a scatter plot of the
pulse heights observed in the
light detector versus the pulse height observed in the phonon detector.  
The pulse height has been converted to 
energy using the 122 keV photon peak in both detectors. 
The plot shows two well-separated bands.
The lower band is caused by neutron-induced nuclear recoils
while the diagonal band is caused by electron recoils induced
by $\gamma$'s and $\beta$'s.
Electron and nuclear recoils can be clearly distinguished down to a
threshold of about 10keV.
From the ratio of  pulse height  light / pulse height phonons   
for the two bands, 
a quenching factor 
(the ratio of the light output for electrons to that for nuclear recoil) 
of 7.4 can be inferred. The
leakage of  electron recoils  into the nuclear recoil band
 determines the effectiveness of the
electron recoil rejection. If we use a quality factor  as defined 
in Ref.~\cite{gaitskell}, 
a detailed evaluation together with data taken without neutrons, 
we find a rejection factor of
98\,\% in the energy
range between 10\,keV and 20\,keV; 99.7\,\% in the range between
15\,keV and 
25\,keV; and better than 99.9\,\% above 20\,keV. 

The simultaneous light and phonon measurement has several
advantages over the simultaneous 
measurement of charge and phonons, which is another method of
the same type. In the measurement of charge and phonons electrical
contacts
always produce a dead layer on the  surface, which
causes surface events, especially electrons from outside, to leak
into the
nuclear recoil band. As our measurements with electrons clearly
show, this problem does 
not exist with light detection. 
Also  light collection does not
suffer from 
problems such as space charge build up, field inhomogeneities or  
phonons produced by drifting the charges. 
Thus many  effects are absent 
which in  charge/phonon measurement can cause leakage of electron recoils
into the nuclear recoil band. As a result the background
suppression efficiency of the light/phonon 
detection is excellent.  It works equally well for photons and
electrons, thus avoiding particle dependent systematic
uncertainties in discrimination.
Furthermore the large quenching factor of the CaWO$_4$ gives a very
effective separation 
of nuclear recoils from electron recoils. 

The possibility of using 
different scintillators with different target nuclei, which is
possible with CRESST technology,  gives a powerful handle for
understanding  and reducing backgrounds. Even the neutron
background, always considered to be the ultimate limitation for
such systems, could be understood by varying the target nuclei.

For a dark matter search, detector masses of several hundred
grams per channel are needed. 
It is therefore necessary to scale up the 6\,g detector used in the
above tests  without loss of performance. 
Since the device consists of two CRESST-type detectors, we are
confident of being able to produce much larger detectors of 
similar performance by applying
familiar techniques and optimizing the design. 

For Phase II of CRESST we plan to install scintillating calorimeters
with a total mass of about 10\,kg, consisting of 33 modules
of 300\,g CaWO$_4$ crystals. 
This assembly will fit easily in the present cold box.
However the number of  SQUID readout channels has to be
upgraded to 66. 
Additionally we plan to install an external muon veto and a
passive neutron shield.
These upgrades are being implemented during 2001
as part of our  move from Hall~B to Hall~A of Gran Sasso.

\subsection{Expected dark matter sensitivity}

\begin{figure}[btp]
\begin{center}
\mbox{
\epsfig{file=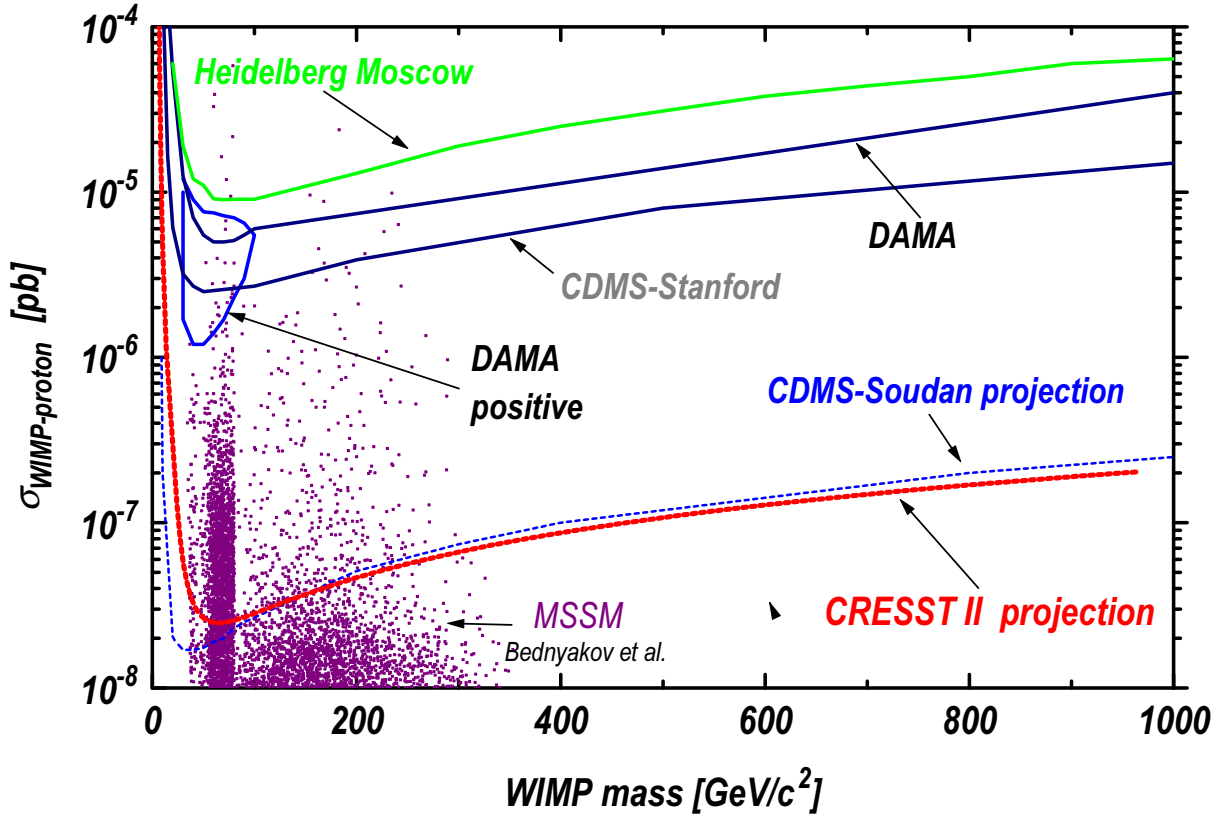,height=16cm,angle=-90}
}
\end{center}
\caption[]{WIMP-nucleon cross section limits (90\,\% CL) for spin-independent
interactions as a function of the WIMP mass, 
expected for a 10\,kg CaWO$_4$
detector with a background rejection of 99.7\,\% above a threshold of 15 keV
detector and  3 years of measurement time in the CRESST set-up in Gran Sasso.
For comparison the measured limit from the Heidelberg-Moscow $^{76}$Ge
experiment~\cite{hdmnew}, 
the DAMA NaI limits~\cite{rita99} 
and contour for positive evidence~\cite{ritapositive}, 
the CDMS Stanford limit~\cite{cdms2000} and the expectation 
for CDMS Soudan~\cite{nam} are also shown. 
The scatter plot shows the 
expectations for WIMP-neutralinos calculated in the MSSM framework with 
non-universal scalar mass unification~\cite{bednyakov}.
}
\label{fig-caw}
\end{figure}

The presence of the heavy tungsten nuclei  make the new detectors
particularly sensitive to a spin-independent interaction of WIMPs,
for which the cross section profits from a large coherence factor
of the order $A^2$, where $A$ is the number of nucleons.
Combined with the strong background rejection,
this gives good sensitivity down to low WIMP cross sections.
Fig.~\ref{fig-caw} shows the expected  sensitivity of Phase II, 
based on a background rate of 1\,count/(kg\,keV\,day),  
an intrinsic background rejection of 99.7\,\%
above a recoil threshold of 15\,keV, and an exposure of 30 kg years.
For comparison the measured limits from the
Heidelberg-Moscow $^{76}$Ge-diode  experiment~\cite{hdmnew}, 
the  DAMA NaI experiment~\cite{rita99}, 
the CDMS experiment at Stanford~\cite{cdms2000}, 
and the projected sensitivity of CDMS at the Soudan mine~\cite{nam} 
are also shown together  with  
the contour for positive evidence~\cite{ritapositive} 
from the DAMA experiment. 

In a 10\,kg CaWO$_4$ detector, 60\,GeV WIMPs with the
cross section claimed in Ref.~\cite{ritapositive}
would give about 46 counts between
15 and 25\,keV within one month. A background of
1~count/(kg\,keV\,day) suppressed with 99.7\,\% 
would leave 9 background counts in the same energy range.
Thus a 10\,kg CaWO$_4$ detector should allow a  test of  the  reported
positive signal with 1 month of measuring time.  

\bigskip
This work was supported by 
the DFG SFB 375 ``Particle Astrophysics'',
the EU Network ``Cryogenic Detectors'' (contract ERBFMRXCT980167),
BMBF, PPARC, 
and two EU Marie Curie Fellowships.

\end{document}